\documentclass[pra,aps,showpacs,showkeys,superscriptaddress,reprint,amsmath,amssymb, footinbib]{revtex4-1}
\usepackage{graphicx}
\usepackage{color}
\usepackage{amssymb,amsmath,graphicx}
\usepackage{hyperref}
\hypersetup{
bookmarks=false,
anchorcolor=blue,
colorlinks=true,
citecolor=blue,
urlcolor=blue,
linkcolor=blue}

\begin{document}

\title{Continuous matrix product states solution for
\\
 the mixing/demixing transition 
in one-dimensional quantum fields}

\author{Fernando Quijandr\'{\i}a}
\affiliation{Instituto de Ciencia de Materiales de Arag\'on y
  Departamento de F\'{\i}sica de la Materia Condensada,
  CSIC-Universidad de Zaragoza, Zaragoza, E-50012, Spain.}
\author{David Zueco} 
\affiliation{Instituto de Ciencia de Materiales de Arag\'on y Departamento de F\'{\i}sica de la Materia Condensada, CSIC-Universidad de Zaragoza, Zaragoza, E-50012, Spain.}
\affiliation{Fundaci\'on ARAID, Paseo Mar\'{\i}a Agust\'{\i}n 36, Zaragoza 50004, Spain}

\date{\today}

\begin{abstract}
We  solve the 
\emph {mixing-demixing transition}
 in repulsive one-dimensional 
\emph {bose-bose mixtures}.
This is done numerically by means of the continuous matrix product states variational \emph{ansatz}.
We show that the effective low-energy bosonization theory is able to detect the transition whenever the Luttinger parameters are \emph{exactly} computed. 
We further characterize the transition by calculating the ground-state energy density, the field-field fluctuations and the density correlations.
\end{abstract}


\maketitle 


\section{Introduction}

Understanding the  physics of strongly correlated  many-body systems is a formidable task, both in the lattice and in the continuum \cite{Cirac2012}. 
There is a fruitful synergy between condensed matter, high energy physics  or quantum chemistry  and  the
quantum information community.
Ideas 
 as tensor network states \cite{Orus2014} or quantum simulations \cite{Georgescu2013}  are pursuing the goal of understanding phases and dynamics beyond the paradigm of perturbative theories.

One-dimensional many-body systems are a good example of this cooperation.  
Well established theoretical techniques as bosonization \cite{Giamarchi}  are complemented with the density matrix renormalization group (DMRG) and matrix product states (MPS) \cite{White1992,Schollwock2011}  in the lattice and more recently, continuous matrix product states (cMPS)  in the continuum \cite{Verstraete2010}. 
Ultracold gases  are a paradigmatic example of experiments realizing one dimensional quantum fields \cite{Bloch2008, Cazalilla2011, Guan2013}.
Experiments and simulations in one dimension are perfect testbeds since each of them can be used for benchmarking the other \cite{Moritz2003,Kinoshita2004,Paredes2004, Trotzky2012}.

Of special relevance for this work is the cMPS formalism.
Introduced by Verstraete and Cirac \cite{Verstraete2010}, these states constitute a \emph{variational class} for the efficient simulation of quantum field theories that does not rely on a space \emph{discretization}.
The ansatz has  proven  to be efficient for computing the ground state, dispersion relation \cite{Draxler2013} and quantum evolution \cite{Haegeman2015} of nonrelativistic theories. In addition, introducing a suitable regularization prescription, it has also been applied to the study of certain relativistic phenomena \cite{Haegeman2013}. 
Finally, the cMPS \emph{ansatz} has been already tested for bose-bose \cite{Quijandria2014} and fermi-fermi \cite{Chung2015} mixtures, both in nonrelativistic setups.

Among the different scenarios covered in experiments, we are interested here in bose-bose mixtures \cite{Modugno2002, Schweigler2015} .
In this work, 
we aim to characterize  the mixing/demixing phase transition occurring in repulsive bosonic mixtures, via cMPS.
Here, the competition between the intra and interspecies couplings leads to the formation of two different phases.  The miscible, were the two gases coexist, and the immiscible, where the two gases separate from each other \cite{Papp2008}.
This transition has been studied analytically within different approaches \cite{Timmermans1998, Cazalilla2003, Kolezhuk2010} but its numerical simulation has been elusive.   

We will review the phase transition, clarify some issues on the instability occurring  already in the Luttinger liquid description, and fully characterize it via the ground state energy, fluctuations and correlation functions.


\section{The phase transition}\label{sect:transition}

Two 1D bosonic gases interacting via a quartic contact potential 
($\hbar=2m=1$) are described in second quantization via the Hamiltonian
\begin{align}
\label{Hamiltonian}
\hat{H} &= \sum_{\alpha = 1}^2 \int_0^L {\rm d}x\,  \partial_x \hat{\psi}^{\dagger}_{\alpha}(x) \partial_x \hat{\psi}_{\alpha}(x) \nonumber\\ &+ \sum_{\alpha,\beta = 1}^2 g_{\alpha \beta} \int_0^L {\rm d}x \, \hat{\psi}^{\dagger}_{\alpha}(x) \hat{\psi}^{\dagger}_{\beta}(x) \hat{\psi}_{\beta}(x) \hat{\psi}_{\alpha}(x) 
\end{align}
where $\hat{\psi}^{\dagger}_{\alpha}(x)$ ($\hat{\psi}_{\alpha}(x)$) are the bosonic field operators which create (annihilate) bosonic particles of species $\alpha$ at the position $x \in [0,L]$. They satisfy the commutation relations: $[\hat{\psi}_{\alpha}(x), \hat{\psi}^{\dagger}_{\beta}(x')]=\delta_{\alpha \beta}\delta(x-x')$ and $[\hat{\psi}_{\alpha}(x), \hat{\psi}_{\beta}(x')]=
[\hat{\psi}_{\alpha}^\dagger(x), \hat{\psi}_{\beta}^\dagger (x')]=
0$. 
In this work, we want to characterize numerically the mixing-demixing transition occurring in mixtures whenever two different repulsive bosonic species are trapped  together.  
We will consider the case where the participating species are \emph{nonconvertible}, {\it i.e.} 
the individual particle densities $\rho_{0\alpha}$ of each bosonic species are conserved separately
 [$\rho_{0\alpha} = \langle \hat{\psi}^{\dagger}_{\alpha}(x) \hat{\psi}_{\alpha}(x) \rangle$,  $ \langle \; \rangle $ means averages over the ground state of \eqref{Hamiltonian}].
We will also restrict to the symmetric case $\rho_{01} = \rho_{02} = \rho_0$,  
$g_{11}=g_{22}=c >0$  and $g_{12}=g_{21}=g/2 >0$.

The mixing/demixing transition has been broadly studied analytically, see \emph{ e.g.} Refs. \onlinecite{Larsen1963, Goldstein1997, Timmermans1998, Ao1998, Kolezhuk2010, Cazalilla2003}.
The phase separation,  which lies on the  competition between the repulsion strengths $c$ and $g$, can be understood on several grounds.
The simplest approach considers a mean-field treatment. Here, the interaction term can be seen as a quadratic form of the densities. The latter is positive defined as long as $g < 2c$. When the positivity condition is violated ($g \geq 2c$) an instability occurs.
In more than one dimension, both species must separate in order to make the overlap integral zero, \emph{i.e.}, minimizing the repulsive interaction which yields the instability.
In this phase  ($g \geq 2c$) both species are immiscible  as observed by resolving the spatial density profiles in the trap  \cite{Papp2008}.
In 1D, there is no possibility of spatial separation. The interspecies fluctuations, $\hat{\psi}^{\dagger}_{1}(x) \hat{\psi}^{\dagger}_{2}(x) \hat{\psi}_{1}(x) \hat{\psi}_{2}(x)$ in Eq. (\ref{Hamiltonian}) are zero in the ground state.

One-dimensional systems are somehow special.  Their confinement provides an enhancement of the collective behaviour, leading to a universality class in the low-energy or long wavelength sector.
The latter is known as Luttinger Liquid
\cite{Haldane1,Haldane2,Giamarchi}.
This regime is described by introducing 
the bosonic operators $\hat{\phi}_{\alpha}$ and $\hat{\theta}_{\alpha}$ in terms of which we rewrite the field operators 
$\hat{\psi}_{\alpha}(x)=(\rho_{0 \alpha} - \partial_x \hat{\phi}_{\alpha}(x) /\pi)^{1/2} \sum_{p=-\infty}^{+\infty}{\rm e}^{ip(\pi \rho_{0 \alpha} x - \hat{\phi}_{\alpha}(x))}{\rm e}^{i\hat{\theta}(x)}$. 
This is nothing but the \emph{harmonic fluid approach} treatment best known in the literature as \emph{bosonization} \cite{Cazalilla2004}.
Note that for high enough values of $p$, the exponential terms oscillate very fast and rapidly average to zero. Therefore, in order to obtain the low-energy \emph{effective} Hamiltonian, we should only keep a few relevant terms. This leads to 
\begin{align}\label{Heff}
2 \pi \hat{H}_{\rm eff}  &= \int {\rm d}x\, \sum_{\alpha = 1}^2 \left( \frac{v_{\alpha}}{K_{\alpha}}(\partial_x \hat{\phi}_{\alpha})^2 + v_{\alpha} K_{\alpha} (\partial_x \hat{\theta}_{\alpha})^2 \right) 
\\\nonumber
 &+ \int {\rm d}x\, \bigg ( 2g_x \partial_x \hat{\phi}_1 \partial_x \hat{\phi}_2 + g_c \cos \Big (2(\hat{\phi}_1 - \hat{\phi}_2) \Big ) \bigg )
\end{align}
This long wavelength  description is fully characterized by the dimensionless parameters $K_{\alpha}$, the velocities $v_{\alpha}$ and the coupling strengths $g_x$ and $g_c$ (Luttinger parameters). For the symmetric case considered here, we have that $v_1 = v_2 = v$ and $K_1 = K_2 = K$. This model can be easily decoupled by introducing the normal modes $\hat{\phi}_\pm = 1/\sqrt{2}(\hat{\phi}_1 \pm \hat{\phi}_2)$ and $\hat{\theta}_\pm = 1/\sqrt{2}(\hat{\theta}_1 \pm \hat{\theta}_2)$. In terms of them, the low-energy Hamiltonian reads
$
2 \pi \hat{H}_{\rm eff} = \int {\rm d}x\, \sum_{\nu = \pm} \left( \frac{v_{\nu}}{K_{\nu}}(\partial_x \hat{\phi}_{\nu})^2 + v_{\nu} K_{\nu} (\partial_x \hat{\theta}_{\nu})^2 \right) + g_c \int {\rm d}x\, \cos(\sqrt{8}\hat{\phi}_-)
$. The normal modes' velocities $v_\pm$ are defined as follows
\begin{equation}\label{v2normal}
v_\pm^2 = 1 \pm \frac{K g_x}{v}
\end{equation}

As pointed out by Cazalilla and Ho in Ref. \onlinecite{Cazalilla2003}, the coupled system \eqref{Heff}  is unstable when $v_-^2$ becomes negative.  In other words, the action is not anymore definite positive, pretty much like in the mean-field argument sketched before.
This will happen whenever $K g_x > v$.  Thus, to compute the transition point, we just need to find the Luttinger parameters from the original Hamiltonian \eqref{Hamiltonian}.
In Ref. \onlinecite{Cazalilla2003}, $K, g_x$ and $v$ were  approximated via expressions valid in the weak interspecies coupling $g$ regime. In the quasi-condensate regime $\gamma = c/\rho \lesssim 1$ \cite{Cazalilla2004}, the instability is estimated to happen at $g_* = 2c(1 - \sqrt{\gamma}/2\pi)$.
We stress that this result deviates from the mean field value $g_* = 2 c$.

The phase separation has also been studied analytically beyond perturbation theory by Kolezhuk \cite{Kolezhuk2010}. He found that for one and two-dimensional gases, the transition point, in the symmetric case, does not depend on the particle densities. Surprisingly enough, the non-perturbative result coincides with the mean-field description. That is, the two species demix when $g \geq g_* =  2c$. 
Following this result, one might be tempted to think that the bosonization framework is not able to predict correctly the transition point. It could be argued that the Luttinger liquid paradigm breaks down at intermediates values of $g$, below the critical value $g_* = 2 c$. 
Here, we will show that this is not the case.  We demonstrate that the bosonization predicts the transition correctly when the Luttinger parameters are computed exactly instead of  using approximations. 


\section{cMPS solution}
A translational invariant cMPS of $N$ species ($N=2$ in this work) of bosonic particles is defined by the state vector \cite{Haegeman2013}:
\begin{align}
\label{cMPS}
\vert \chi \rangle = {\rm Tr}_{\rm aux} \mathcal{P} {\rm exp} \left( \int_0^L {\rm d}x \, \widetilde Q \otimes \mathbb{I} + \sum_{\alpha=1}^2 
\widetilde R_{\alpha} \otimes \hat{\psi}^{\dagger}_{\alpha}(x) \right) \vert \Omega \rangle
\end{align}
where
$\hat{\psi}_{\alpha}(x)$ are the bosonic field operators, $\widetilde Q$ and $ \widetilde R_{\alpha}$ are a set of complex,
 $\widetilde D \times \widetilde D$ matrices acting on an \emph{auxiliary} 
$\widetilde D$-dimensional space and $\vert \Omega \rangle$ is the \emph{free} vacuum state vector ($\hat{\psi}_\alpha (x) |\Omega\rangle =0$). $\mathcal{P}$ denotes a path-ordering prescription and the partial trace, ${\rm Tr_{\rm aux}}$,  is taken over the auxiliary space. 
This way of writing field states is the continuous limit of a MPS \cite{Verstraete2010}. 
 Here, the dimension $\widetilde D$ of the auxiliary matrices corresponds to  
the so-called \emph{bond dimension},  an upper bound to the entanglement entropy. Typically, the low-energy states of local Hamiltonians should possess a low amount of entanglement,  consequently $\widetilde D$ is a \emph {small} number. 
Being the bond dimension small, the state \eqref{cMPS} represents an efficient trial for finding the ground state of one-dimensional field theories numerically.

In a previous work \cite{Quijandria2014} the authors showed how to construct a two-species cMPS starting from two decoupled single species solutions. In brief, for coupled fields we considered coupled auxiliary spaces (one per bosonic field).   The total auxiliary Hamiltonian was extended to ($\widetilde K = -i \widetilde Q - \frac{1}{2} \sum \widetilde R_\alpha^\dagger \widetilde R_\alpha$)
\begin{equation}
\label{K}
\widetilde K  = K_1 \otimes {\mathbb I}_2 + {\mathbb I}_1 \otimes K_2 
+
\sum_{p=1}^P Z_1^{(p)} \otimes Z_2^{(p)}
\; .
\end{equation}
where $K_{\alpha}$ is the auxiliary Hamiltonian associated to bosonic species $\alpha$. The parameter $P$ accounts for the number of pairs of coupling matrices entering in the cMPS state. Consequently, the matrices $\widetilde R_{\alpha}$, belonging to the auxiliary space of field $\alpha$, were extended into the total product space: $\widetilde R_1 = R_1 \otimes \mathbb{I}$ and $\widetilde R_2 = \mathbb{I} \otimes R_2$.  
Denoting $D$ the dimension of the matrices $R_1$ and $R_2$ the bond dimension is  then  $\widetilde D = D^2$.
The total number of variatonal parameters is $D^2 (4 + 2 P )$.
 Details can be found in Ref. \onlinecite{Quijandria2014}.

In the thermodynamic limit ($L \to \infty$), the fluctuations and correlation functions can be computed from
\begin{align}
 \label{C}
C_{\alpha \beta}(x-y) &\equiv \langle \hat{\psi}^{\dagger}_{\alpha}(x) \hat{\psi}^{\dagger}_{\beta}(y) \hat{\psi}_{\beta}(y) \hat{\psi}_{\alpha}(x) \rangle
\\ \nonumber 
&= {\rm Tr}[( \widetilde R_{\beta} \otimes \widetilde R_{\beta}^*)
 {\rm e}^{T (x-y)} 
(\widetilde R_{\alpha} \otimes \widetilde R_{\alpha}^*)]
\; , 
\end{align}
without loss of generality, we have assumed that $x>y$. 
Remind that,  throughout this work $\langle \; \rangle$ means averages over the ground state of \eqref{Hamiltonian}. The transfer operator $T$ is defined as: $T \equiv \widetilde Q \otimes \mathbb{I} + \mathbb{I} \otimes \widetilde Q^* + \sum_{\alpha=1}^2 \widetilde R_{\alpha} \otimes \widetilde R_{\alpha}^* $. Finally, note that the fluctuations are calculated by making $x=y$ in \eqref{C}.


\begin{figure}
\begin{center}
\includegraphics[width=0.99\columnwidth]{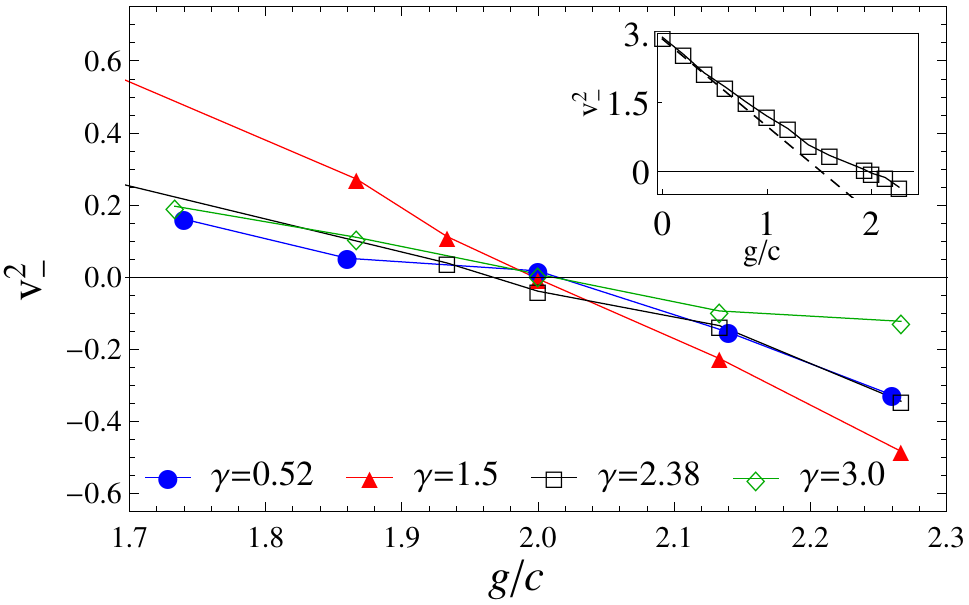}
\end{center}
\caption{
(Color online) Instability in the bosonization description. The square velocities $v_-^2$,  defined by \eqref{vpm}, are calculated using cMPS for different values of the parameter $\gamma = c/\rho$: $0.52$ (filled circles), $1.5$ (filled triangles), $2.38$ (open squares) and $3.0$ (open diamonds). In the inset, we compare the numerical result for $\gamma = 2.38$ (open squares) with a weak-coupling estimation (dashed line) for the same value of $\gamma$ with $c=1.5$ and $\rho=0.63$ (see the main text). All of the simulations have been performed with $D=5$ and $P=1$.
}
\label{fig:square-minus}
\end{figure}

\section{Results}

As already anticipated, our goal is to characterize the mixing/demixing transition numerically. 
We do it in two ways.  First, we study the instability in the low-energy regime described by the effective Hamiltonian \eqref{Heff}.  The second strategy  is to look directly at the ground state of \eqref{Hamiltonian} and  compute the fluctuations and correlation functions, Cf. Eq. \eqref{C}.


\subsection{Bosonization instability}

In the harmonic fluid approach the normal modes for the fields decouple (see the discussion below Eq. \eqref{Heff}). 
Each of these modes propagate with different velocities, $v_\pm$.
Within the bosonization framework, these velocities can be related to the  ground state energy density ($e_0$). The explicit expressions for the velocities are \cite{Cazalilla2004, Kleine2008} 
\begin{equation}\label{vpm}
v_\pm^2 = 2 \rho_\pm \frac{\partial^2 e_0}{\partial \rho_\pm^2}
\end{equation}
with $\rho_\pm = \rho_1 \pm \rho_2 $.
Analytical estimations for these velocities follow from Eq. \eqref{v2normal}. In the weak-coupling regime ($g \ll c$), it is safe to assume that $v$ and $K$ correspond to the solutions for a single bosonic field \cite{Cazalilla2004}. In turn, $g_x$ is approximated by $g_x \simeq g/\pi$ (see reference \onlinecite{Cazalilla2003}).

In the inset of figure \ref{fig:square-minus}, it can be seen that 
already at intermediate values of $g$ (well below the critical value $g_*$) the predicted velocities $v_\pm$ using weak-coupling analytical expressions deviate from the numerically computed ones \cite{Kleine2008, Quijandria2014}.  
A consequence of this deviation is the failure on the estimation of the point where $v_-^2$ becomes negative, which in turns marks the critical value $g_*$.
In the main figure \ref{fig:square-minus}  we have zoomed the $v_-^2$ around the transition point for different values of $\gamma = c/\rho$.
As it has been already pointed out, within the weak-coupling treatment, the transition is estimated to happen at
$g_*=2c ( 1 - \sqrt{\gamma}/2 \pi )$. As $\gamma = c/\rho$, the latter result makes the transition point dependent on both, the intraspecies coupling $c$ and the particle density $\rho$. 
On the other hand, once $v_-$ is exactly derived from the ground-state energy density by using relation \eqref{vpm}, we see how the transition point becomes independent of $\gamma$. In fact, we see that the mode propagating with velocity $v_-$ becomes ill-defined at $g_*/c = 2$
\cite{Note-Error}, in agreement with 
the mean-field and Kolezhuk results \cite{Kolezhuk2010}.
Therefore, once the Luttinger parameters are exactly computed, the bosonization predicts correctly the transition.  
%


\subsection{Characterization beyond bosonization}

\begin{figure}[t]
\includegraphics[width=0.99\columnwidth]{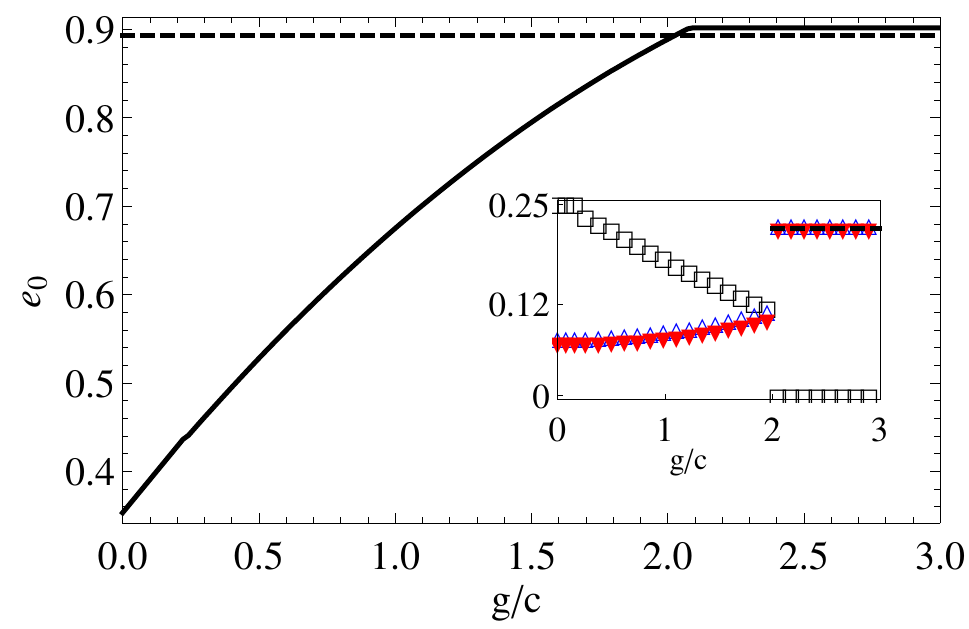}
 \caption{(Color online) Ground state energy density of Hamiltonian \eqref{Hamiltonian} as a function of the coupling ratio $g/c$ calculated with cMPS. We keep fixed the intraspecies coupling $c=1.5$ while the particle density in each of the gases is equal to $\rho =0.5$. In the inset we show the ground state fluctuations as a function of $g/c$: $C_{11}(0)$ (open triangles), $C_{22}(0)$ (inverted filled triangles) and $C_{12}(0)$ (open squares) defined in \eqref{C}. Simulations have been performed with $D=5$ and $P=1$.}
 \label{fig:energy}
 \end{figure}

Having a full knowledge of the ground state of \eqref{Hamiltonian}, we proceed now to 
characterize the phase transition beyond the bosonization formalism.
In figure \ref{fig:energy}, we see the behaviour  of the ground state energy density as a function of the interspecies coupling.  It is direct to realize that after $g_*/c= 2$ the energy remains constant. In this region, 
the ground state is such that the last term of \eqref{Hamiltonian}, \emph{i.e.}, the one accounting for the interaction among different fields, has zero average.  In other words, after the transition we have that:  $C_{1,2}(0) = \langle \hat{\psi}^\dagger_1 (x) \hat{\psi}_1 (x) \hat{\psi}^\dagger_2 (x) \hat{\psi}_2(x) \rangle = 0$, which is explicitly represented in the inset of figure \ref{fig:energy} (open squares).  
This confirms our previous exposition for the phase transition: in one dimension, phase separation implies zero interspecies fluctuations.

Apart from the transition point estimation and the zero field-field overlapping nature for the demixed phase, we can go further in characterizing the properties of the ground state before and after the transition.  
Let us start with the mixed phase.
By looking at the inset of figure \ref{fig:energy} 
we see that the fluctuations $C_{1,2}(0)$ do not remain constant as soon as the interaction is switched on.
The latter behaviour reflects a sublinear growth of $e_0$ in terms of $g$. This means that a simple mean field theory 
$\langle \hat{\psi}^\dagger_1 (x) \hat{\psi}_1 (x) \hat{\psi}^\dagger_2 (x) \hat{\psi}_2(x) \rangle \cong  \rho_1 \rho_2$ is not sufficient for describing this phase.

We will discuss now  the demixed phase. As explained above, after the transition $C_{1,2}(0) = 0$.
It is straightforward to see that a ground state of the form
\begin{equation}
\label{GSd}
| X_{\rm dm} \rangle
=\frac{1}{\sqrt{2}}
\Big (
 | \chi_{2 \rho} \rangle \otimes | \Omega \rangle
+
{\rm e}^{i \theta} 
| \Omega \rangle \otimes | \chi_{2 \rho} \rangle 
\Big ) 
\end{equation}
fulfils this condition (suffix dm stands for demixing).  
Besides, $| X_{\rm dm} \rangle$ must satisfy the
particle density conservation for each bosonic species:
$\langle X_{\rm dm}  | \hat{\psi}^\dagger_\alpha (x)  \hat{\psi}_\alpha (x)  |  X_{\rm dm}  \rangle = \rho$, which in turn  imposes that: 
$
\langle \chi_{2 \rho}  | \hat{\psi}^\dagger_\alpha (x)  \hat{\psi}_\alpha (x)  |  \chi_{ 2 \rho}  \rangle = 2 \rho
$.
Indeed, this is confirmed in Fig. \ref{fig:energy} via the ground state energy density.
We see that after the
transition, $e_0$ is the energy of a single bosonic gas (Lieb-Liniger model) with self-interaction $c$ but double particle density $2\rho$ (dashed line) \cite{Note-Error}.
Finally, by looking at the fluctuations $C_{\alpha \alpha} (0)$, we check that they coincide with those of a single bosonic gas with self-interaction $c$ and particle density $2 \rho$, divided by a factor of two due to normalization in \eqref{GSd}. The fluctuations of a single gas are shown in the inset of figure \ref{fig:energy} with a dashed line.


\begin{figure}
  \includegraphics[width=0.49\columnwidth]{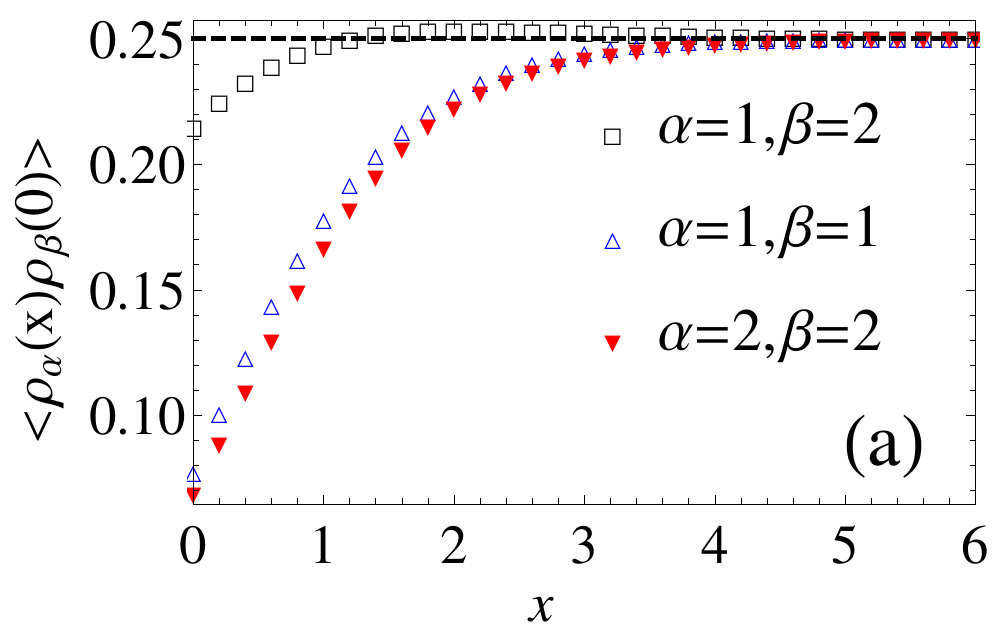}
  \includegraphics[width=.49\columnwidth]{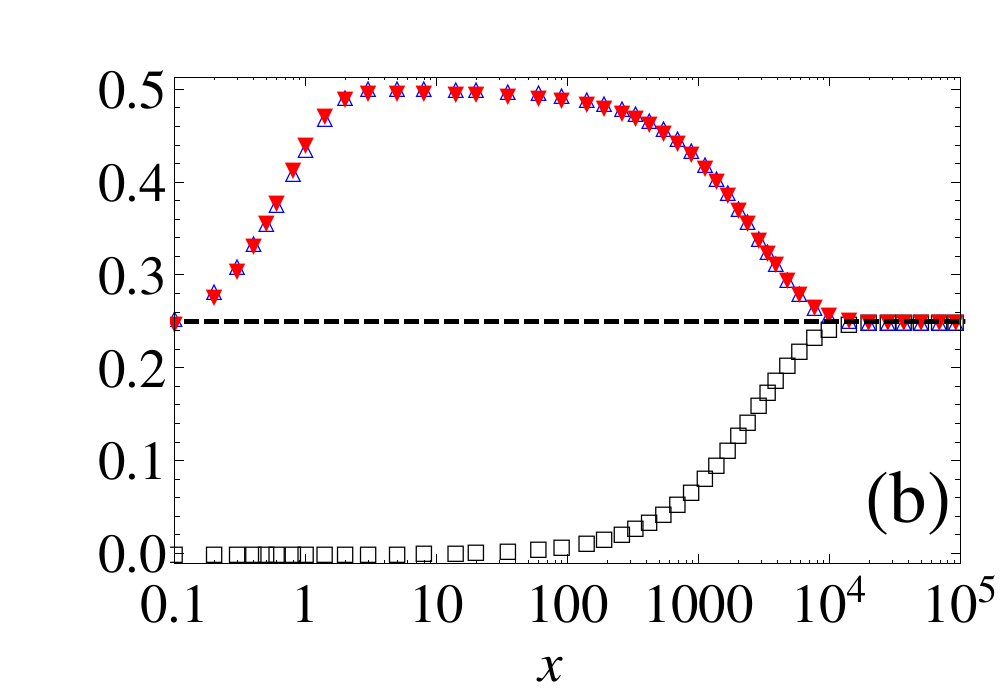}
\caption{(Color online) Density correlation functions: $C_{11}(x)$ (open triangles), $C_{22}(x)$ (inverted filled triangles) and $C_{12}(x)$ (open squares) as a function of the distance $x$ for the same parameters of Fig. \ref{fig:energy}. The transition happens at $g_*/c = 2$. We plot the correlations (a) before the transition $g/c = 0.52$ and (b) after the transition $g/c = 2.53$. The shape of this curve brings to mind the popular story of the boa constrictor digesting an elephant \cite{Exupery}. Simulations have been performed with $D=5$ and $P=1$.}
\label{fig:corr}
\end{figure}

We finish our phase characterization by studying the correlation functions, 
$C_{\alpha \beta} (x)$.
The results are plotted in figure  \ref{fig:corr}.  By definition, the correlations at zero distance match the fluctuations.  On the other hand, in the limit $x\to \infty$, the correlations  factorize yielding $C_{\alpha \beta} (x\to \infty) = \rho^2$ [marked with  dashed lines in \ref{fig:corr} a) and b)].
In the mixed phase  the correlation length is of the order of $x
\cong 5$, pretty much the same than for a single bosonic species with self-interaction $c$ and particle density $\rho$.

More structure  for $C_{\alpha, \beta} (x)$ appears in the demixed phase.  The interspecies correlation function $C_{12} (x)$, obviously starting at zero, has a large correlation length $\sim 10^4$ (notice the logarithmic scale).
To understand this large correlation length we recall that after the
transition the fields are infinitely repelled.
Our interpretation is reinforced by looking at $C_{\alpha \alpha} (x)$.  In the range $0<x<10$ the correlations build up to $2 \rho^2$ wich means that they can be approximated by 
$C_{\alpha \alpha } (x) \cong 1/2 \langle \chi_{2 \rho} | \hat{\rho}_\alpha | \chi_{2 \rho} \rangle^2 = 2 \rho^2$. 
Therefore, the coherence has been lost at the single field level. However, the fully uncorrelated state will involve the full state $|X_{\rm dm} \rangle$ and pretty much  like for the $C_{12} (x)$ correlations, the demixed phase is equivalent to an infinite repulsive phase, explaining again the large coherence lenght to reach  the asymptotic limit $C_{\alpha \alpha} (x\to \infty) = \rho^2$.


\section{Conclusions}\label{conc}

Summarizing, by means of cMPS we have
computed numerically the ground state of two repulsive one-dimensional bosonic nonconvertible fields. This kind of systems exhibits the so-called mixing/demixing phase transition. 
We have validated previous analytical results for the transition point.  Furthermore, we have demonstrated that this point can be resolved within the Luttinger liquid formalism whenever the effective parameters of the theory are calculated exactly.
All this marks a step forward for the cMPS method, here, resolving a phase transition in a non-trivial quantum field theory.

\section*{Acknowledgements}
We acknowledge discussions with Juanjo Garc\' ia-Ripoll and Jorge Alda.
We also acknowledge the support from the Spanish DGICYT under Project No. FIS2011-25167 as well as by the 
Arag\'on (Grupo FENOL) and the EU Project PROMISCE. \\\\

\bibliographystyle{apsrev4-1}
\bibliography{demix}

\begin{thebibliography}{34}
\bibitem [{Not()}]{Note-Error}%
  \BibitemOpen
  \href@noop {} {}\bibinfo {note} {The dispersion in the estimation of $g_*$
  shown in the paper is arround $2 \%$. This agrees with the accuracy in the
  estimation for $e_0$, Cf. Fig. \ref{fig:energy}. Besides, errors can also be
  estimated for the energy of the ground state in the single Lieb Liniger model
  as a function of $D$. Again, for $D=5$, it is of the order of $ 1
  \%$.}\BibitemShut {Stop}%
\end{thebibliography}

\end{document}